\documentclass[aps,prl,twocolumn,showpacs,superscriptaddress]{revtex4-1}
\usepackage{amsmath}
\usepackage{epsfig}
\usepackage{color}
\usepackage{endnotes}
\usepackage{natbib}
\usepackage[colorlinks,breaklinks]{hyperref}
\usepackage{nicefrac}
\usepackage{bm}
\usepackage{dcolumn}
\usepackage{graphics}
\usepackage{graphicx} 
\usepackage{tikz} 
\usepackage{amsmath}
\usepackage{amssymb}
\usepackage{color}
\usepackage{changes}
\usepackage{multirow}
\usepackage{lineno}
\usepackage[normalem]{ulem}
\usepackage{xcolor}

\newcommand{\be}{\begin{equation}}
\newcommand{\ee}{\end{equation}}
\newcommand{\bea}{\begin{align}}
\newcommand{\eea}{\end{align}}
\newcommand{\beqa}{\begin{eqnarray}}
\newcommand{\eeqa}{\end{eqnarray}}

\newcommand*{\MIT }{Massachusetts Institute of Technology, Cambridge, Massachusetts 02139, USA}
\newcommand*{\ODU}{Old Dominion University, Norfolk, Virginia 23529, USA}
\newcommand*{\JLAB}{Thomas Jefferson National Accelerator Facility, Newport News, Virginia 23606, USA}
\newcommand*{\TAU }{School of Physics and Astronomy, Tel Aviv University, Tel Aviv 69978, Israel}
\newcommand*{\Penn}{Pennsylvania State University, University Park, PA, 16802, USA}
\newcommand*{\GW}{George Washington University, Washington, D.C., 20052, USA}

\newcommand*{\UW}{Fermi National Accelerator Laboratory, Batavia, IL 60510, USA}
\begin{document}


\title{Nucleon off-shell structure and the free neutron valence structure from A=3 inclusive electron scattering measurements}

\author{E.P. Segarra}
\affiliation{\MIT}
\author{J.R. Pybus}
\affiliation{\MIT}
\author{F.~Hauenstein}
\affiliation{\MIT}
\affiliation{\ODU}
\author{T. Kutz}
\affiliation{\MIT}
\affiliation{\TAU}
\author{D. Higinbotham}
\affiliation{\JLAB}
\author{G.A. Miller}
\affiliation{\UW}
\author{E. Piasetzky}
\affiliation{\TAU}
\author{A. Schmidt}
\affiliation{\GW}
\author{M. Strikman}
\affiliation{\Penn}
\author{L.B. Weinstein}
\affiliation{\ODU}
\author{O. Hen}
\affiliation{\MIT}
\email[Contact Author \ ]{(hen@mit.edu)}

\begin{abstract}
  Understanding the differences between the distribution of quarks
  bound in protons and neutrons is key for constraining the mechanisms
  of SU(6) spin-flavor symmetry breaking in Quantum Chromodynamics (QCD).  
  While vast amounts of proton structure measurements were done,
  data on the structure of the neutron is much more spars
  as experiments typically extract the structure of neutrons from measurements
  of light atomic nuclei using model-dependent corrections for nuclear effects.
  Recently the MARATHON collaboration performed such an extraction by measuring
  inclusive deep-inelastic electron-scattering on helium-3 and tritium mirror nuclei where 
  nuclear effects are expected to be similar and thus be suppressed in the helium-3 to tritium ratio. 
  Here we evaluate the model dependence of this extraction by examining a wide
  range of models including the effect of using instant-form and light-cone nuclear
  wave functions and several different parameterizations of nucleon modification effects, 
  including those with and without isospin dependence.
  We find that, while the data cannot differentiate among
  the different models of nuclear structure and nucleon modification, they
  consistently prefer a neutron-to-proton structure function ratio of at $x_B \rightarrow 1$
  of $\sim 0.4$ with a typical uncertainty ($1\sigma$) of $\sim0.05$ and $\sim0.10$ for isospin-independent 
  and isospin-dependent modification models, respectively.
  While strongly favoring SU(6) symmetry breaking models based on perturbative QCD and the Schwinger–Dyson equation calculation,
  the MARATHON data do not completely rule out the scalar di-quark models if an isospin-dependent modification exist.
\end{abstract}

\maketitle

Detailed studies of the distribution of quarks bound in nucleons provides crucial insight into the theory of Quantum Chromodynamics (QCD), thereby improving our understanding of emergent QCD phenomena
such as baryon structure, masses, and magnetic
moments, the structure of matter, and the origin of visible mass in the universe~\cite{Bashir:2012,Roberts:2013mja}.
A typical measurement used for such studies is inclusive lepton deep-inelastic scattering (DIS)~\cite{Bjorken:1969ja}.

Comparing the distribution of high-momentum quarks in the proton and neutron is especially sensitive to models of SU(6) spin-flavor symmetry breaking mechanism in QCD, and thus have far reaching implications to our understanding of the fundamental structure of matter~\cite{Roberts:2013mja}.
While vast amounts of proton DIS data exist, the lack of a free neutron target prevents equivalent measurements of the
neutron DIS, challenging our ability to perform a direct test of QCD symmetry breaking mechanisms. 
Instead, the free neutron structure is extracted from measurements that are probing protons and
neutrons bound in atomic nuclei, applying model-dependent theoretical corrections for nuclear effects such as binding, nucleon-motion, and nucleon structure modification~\cite{Segarra:2019gbp,Dulat:2015mca,Accardi:2016qay,Arrington:2011qt}. 
The latter are minimized, but not cancelled, by focusing on the lightest atomic nuclei with 2 and three nucleons, where
exact nuclear wave functions can be calculated and the effects of nucleon binding and Fermi motion and nucleon modification are generally small.

Here we use a wide model phase-space to analyze the recent Helium-3 and Tritium mirror-nuclei DIS measurements of the MARATHON collaboration~\cite{Abrams:2021xum} to examine the model systematic uncertainty associated with the constrain in places on the distribution of valance quarks in the free proton and neutron. Significant emphasis is placed on the correlation between the free neutron structure and bound nucleon modification which prevented such extractions using previous Deuterium and Helium-3 measurements but is mitigated by the use of Tritium and assumptions of isospin symmetry.

\begin{figure*}[t]
\includegraphics[width=1\columnwidth]{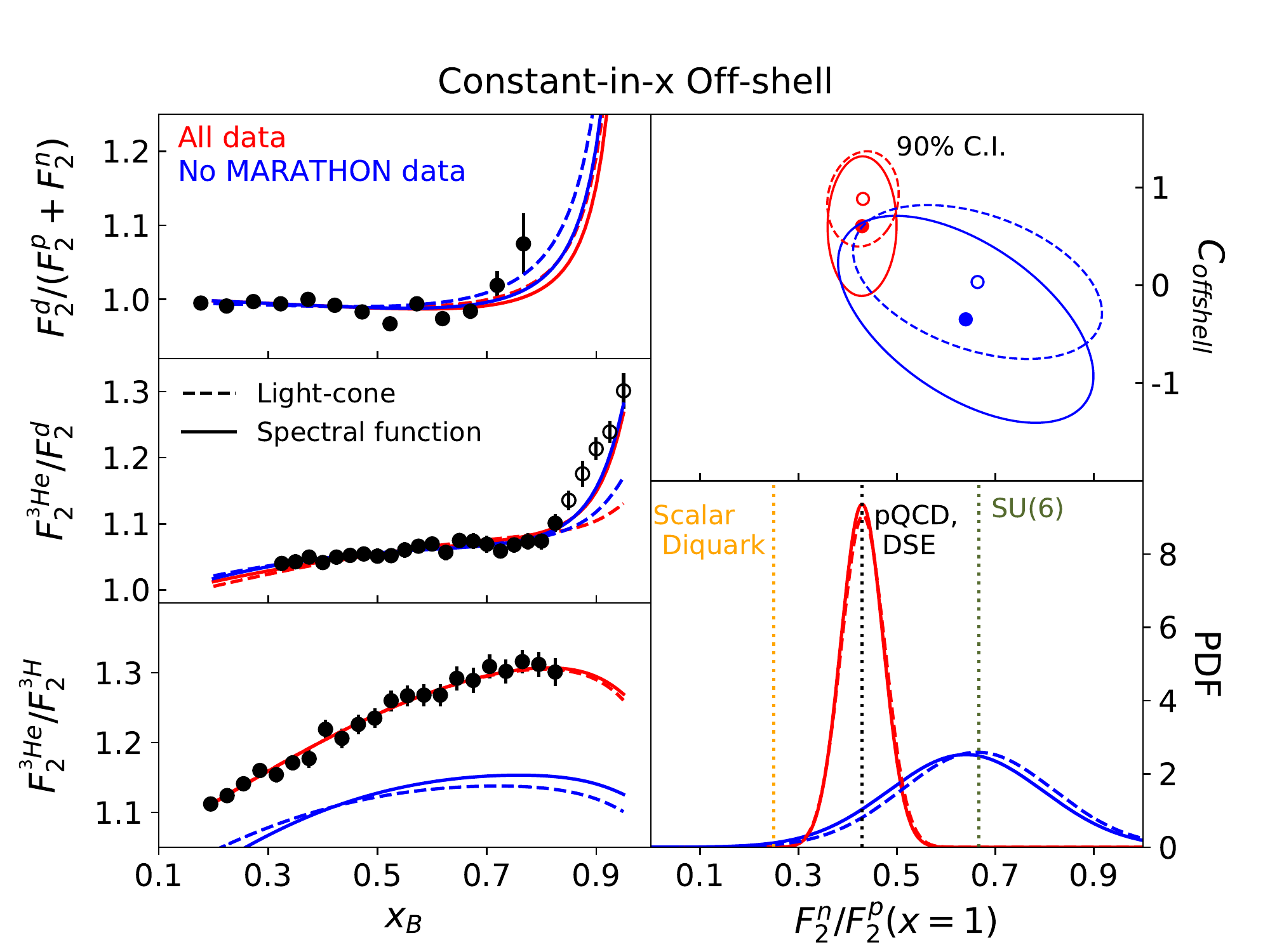}
\includegraphics[width=1\columnwidth]{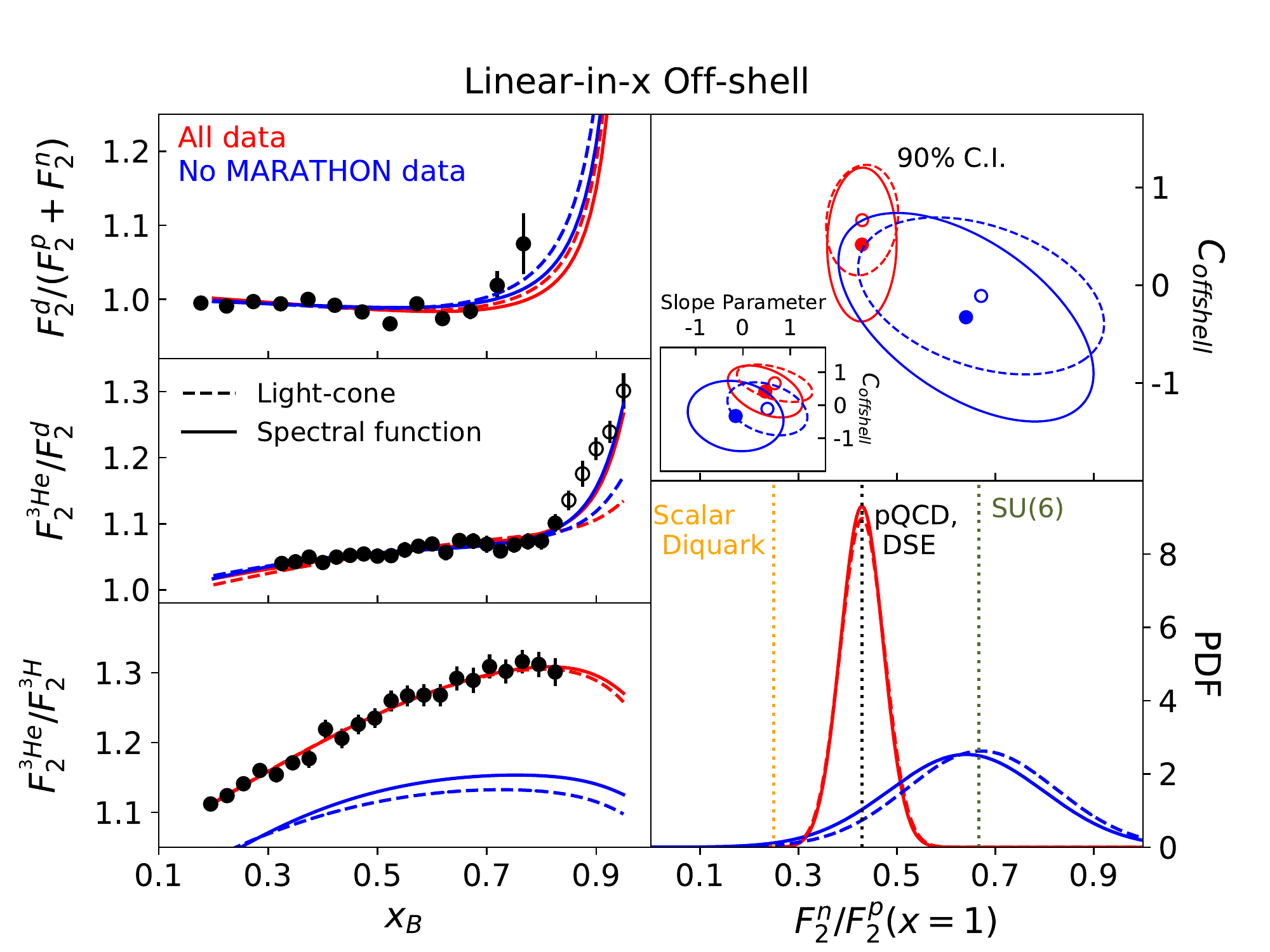}
	\caption{ Fit results for an isospin-independent off-shell nucleon modification function. Left: results for a constant-in-$x$
          off-shell function. Right: results for a linear-in-$x$ off-shell function. In each block, the three left panels show the
          best-fit results for the BONuS (top) \cite{Griffioen:2015hxa}, Seely (middle) \cite{Seely:2009gt}, and MARATHON (bottom) \cite{Abrams:2021xum} data. 
          The upper right panels shows a two-dimensional 90\%
          confidence region for $F_2^n/F_2^p\big|_{x_B\to1}$ and one off-shell parameter. The lower right panels shows the best-fit result
          translated into a probability distribution function (PDF) for the extracted value of $F_2^n/F_2^p\big|_{x_B\to1}$ expected 
          if the fit procedure were repeated on an equivalent independent data set. Also, values of $F_2^n/F_2^p\big|_{x_B\to1}$ predicted
          by various models (translated from $d/u\big|_{x\to1}$ using the simple parton model) are indicated by dotted vertical lines. 
          As described in the text, blue curves show
          results when excluding MARATHON data, while red curves include MARATHON data in the fit. Solid curves use the spectral-function
          formulation for $\rho_N^A(\alpha,v)$, while dashed-curves use the light-cone formulation. 
	}
\label{Fig:res1}
\end{figure*}

The cross-section for lepton DIS off stationary nucleon targets at fixed momentum transfer squared $Q^2=|\vec{q}|^2-\nu^2$ and 
Bjorken scaling parameter $x_B = Q^2/2m_N\nu$ is directly proportional to the inelastic structure function of the target nucleon, 
$F_2(Q^2, x_B)$ in the limit $F_2=2xF_1$.
Here $q=(\vec{q},\nu)$ is the four-momentum transfer from the electron to the target carried by the exchange of a virtual photon, 
and $m_N$ is the nucleon mass. The parameter $x_B$ coincides, in the infinite-momentum frame and $Q^2\rightarrow \infty$,
with the momentum fraction, $x$, of a
parton struck in the scattering reaction. 

In the simple quark-parton model, $F_2(Q^2, x_B)$ is the linear combination of individual quark 
flavor distributions (up, down, strange...), each weighted by the square of its electromagnetic charge.
Therefore, $F_2(Q^2, x_B)$ structure function extractions from lepton-nucleon DIS measurements contain valuable
information on the internal structure of the nucleon.

To test the finer predictions of QCD, it is highly desirable to separate $F_2(Q^2, x_B)$ into the individual quark flavor distributions.
Specifically, the ratio of the proton's down to up quark distributions, $d/u$, in the $x \rightarrow 1$ limit has been 
shown to be extremely sensitive to the SU(6) spin-flavor symmetry breaking mechanism in QCD~\cite{Roberts:2013mja}. 
SU(6) is a broken symmetry in QCD, 
as is clear from the mass difference between the nucleon and $\Delta$-baryon, but if it were preserved, then the proton's $d/u$ 
ratio at $x=1$ would be 1/2~\cite{Close:1979bt}; the proton has two valence up quarks, but only a single valence down quark. Specific models of the 
symmetry breaking mechanism predict a range of end points for $d/u$. A scalar di-quark picture predicts 
$d/u\big|_{x\to1} = 0$~\cite{Close:1973xw,Carlitz:1975bg}, 
perturbative QCD predicts $d/u\big|_{x\to1}= 1/5$~\cite{farrarjackson75}, 
and modern calculation using the Dyson-Schwinger equation predict a range
of end points between 0.18 and 0.27~\cite{Roberts:2013mja}. A precise experimental determination would therefore offer valuable insight into how
this symmetry is broken in QCD.

\begin{figure*}[t]
\begin{center}
\includegraphics[width=1\columnwidth]{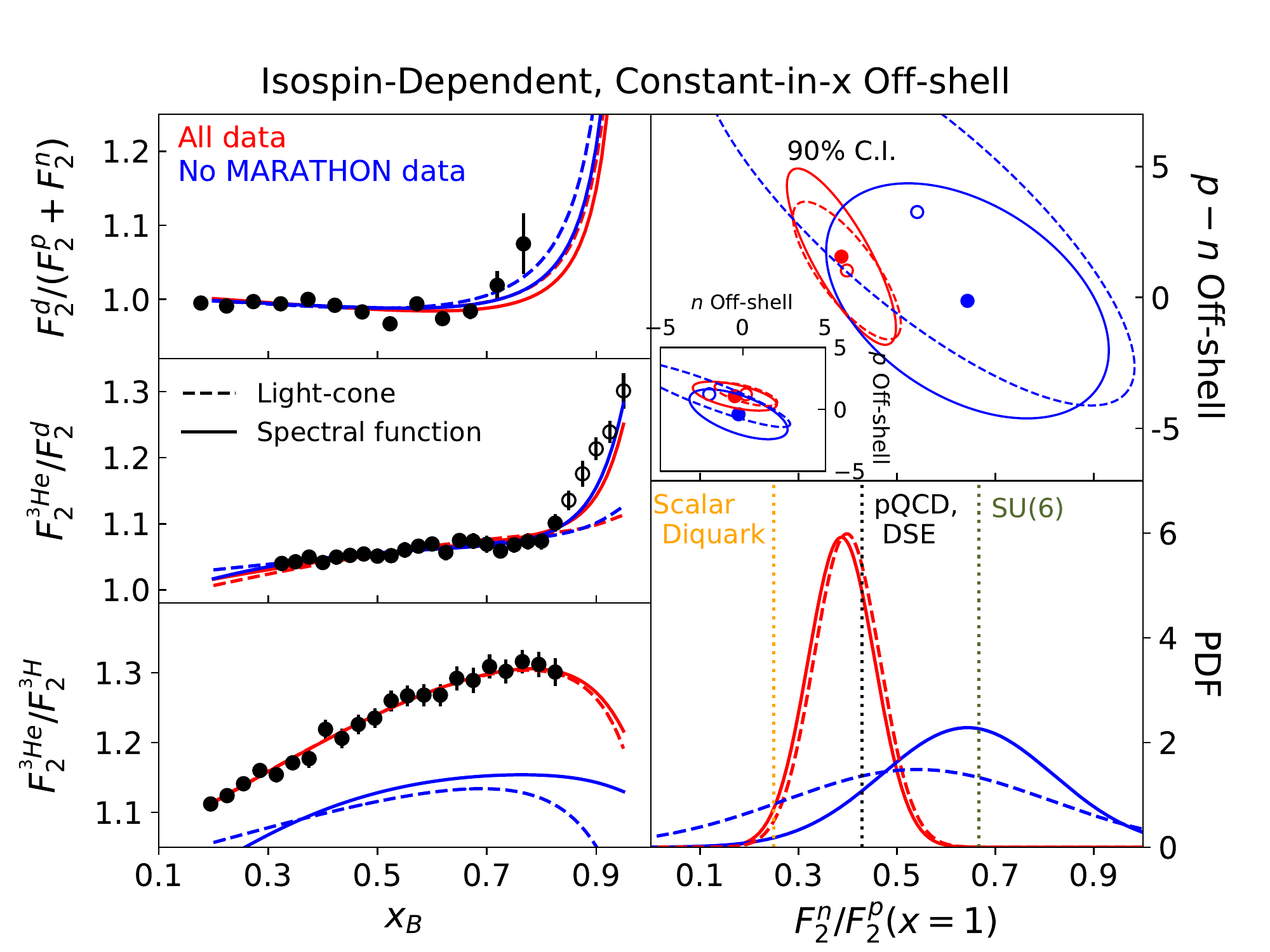}
\includegraphics[width=1\columnwidth]{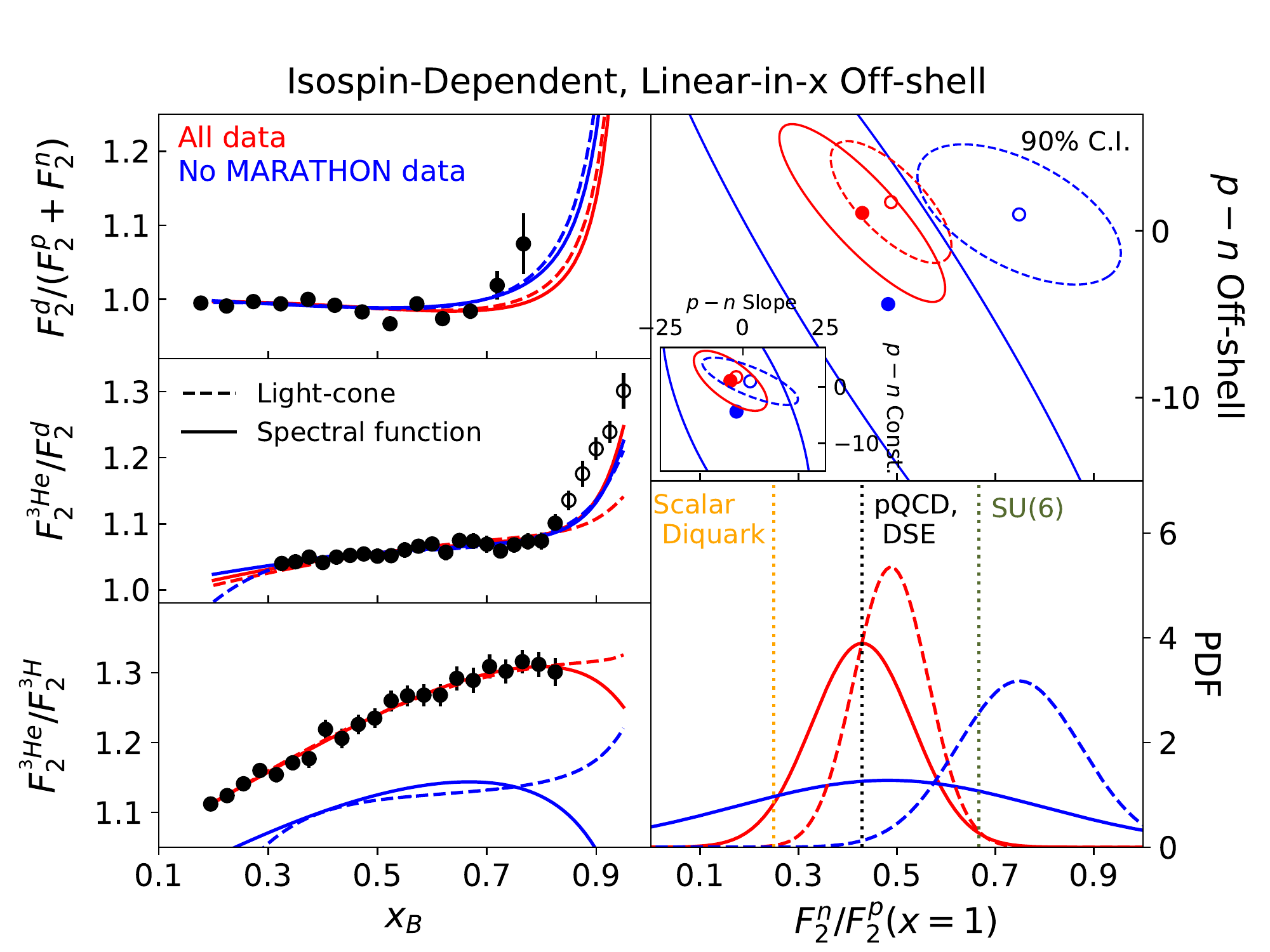}
	\caption{ Fit results for an isospin-dependent off-shell nucleon modification function. The panels
          are arranged as in Fig.~\ref{Fig:res1}. 
	}
\label{Fig:res_iso}
\end{center}
\end{figure*}

Performing such flavor decomposition is a challenging task, which is often met by using isospin symmetry in comparing data from free proton and bound neutron measurements of light-nuclei, with appropriate theoretical corrections to obtain the free neutron structure. 
While advances in nuclear structure theory have greatly improved the accuracy of nucleon motion and binding effects, there is still the problem of the largely unconstrained effects of the modification of the quark distributions in bound nucleons~\cite{Hen:2016kwk}.
This bound structure modification is known as the EMC effect, and is the main limitation in extracting the free neutron
structure function from inclusive DIS measurements of light nuclei.

This fundamental limitation is currently being addressed by several novel measurements, such as those of the BONuS~\cite{Baillie:2011za,tkachenko14,bonus12}
and MARATHON~\cite{MARATHON,Abrams:2021xum} experiments. BONuS measures tagged-DIS (TDIS) reactions to identify scattering events from nearly free neutrons in deuterium. 
MARATHON measures the ratio of the inclusive DIS cross-section of helium-3 ($^3$He) and tritium ($^3$H) mirror 
nuclei, where nuclear effects are expected to mostly cancel in the ratio due to the similarity in the nuclear structure of the two nuclei.
Recently, the MARATHON collaboration published its first results for the $^3$He/$^3$H DIS cross-section ratio and, 
using the calculations of Kulagin and Petti (KP)~\cite{Kulagin:2019pc,Kulagin:2010gd}, applied model-dependent corrections to extract the free neutron to 
proton structure function ratio $F_2^n(Q^2, x_B) / F_2^p(Q^2, x_B)$~\cite{Abrams:2021xum}.

Inspired by previous studies that indicate a potential systematic model uncertainty of up to $20\%$ 
in this extraction~\cite{Segarra:2019gbp,Segarra:2020plg}, 
we examine the impact of the variation across the space of models on the accuracy of the $F_2^n(Q^2, x_B) / F_2^p(Q^2, x_B)$ extraction
from the MARATHON data. Our study includes both instant-form and light-cone nuclear wave functions as well as a broad range of nuclear 
modification models that all reproduce existing DIS data from light nuclei~\cite{Segarra:2020plg}.

Within the standard nuclear convolution approximation, the measured inelastic structure function of a nuclear target,
$F_2^A(x_B)$, is related to the free proton and neutron structure functions 
by~\cite{Frankfurt:1985cv,FSemc:1987,Frankfurt88,SAKulaginSVAku:1985,DunneThomas:1985}:
\begin{small}
\begin{equation}
\begin{split}
		&F_2^A(x_B) = \\
		& \frac{1}{A}\int_{x_B}^A \frac{d\alpha}{\alpha} \int_{-\infty}^0 dv \: F_2^p\big(\tilde{x}\big) \left[ Z \tilde{\rho}_p^A(\alpha,v) + 
		N \tilde{\rho}_n^A(\alpha,v) \frac{F_2^n(\tilde{x})}{F_2^p(\tilde{x})} \right]   \\
		& \;\;\;\;\;\;\;\;\;\;\;\;\;\;\;\;  \times  \Big(1+v\ f^\text{off}(\tilde{x})\Big),
\label{Eq:F2_conv}
\end{split}
\end{equation}
\end{small}
where $Z$ and $N$ are respectively the proton and neutron numbers
of the target nucleus, $A=Z+N$, $\alpha$ is the nucleon light-cone
momentum fraction, $v$ is the fractional virtuality of the bound nucleon defined by $v = (E^2-|{\bf{p}}|^2 - m_N^2)/m_N^2$,
$\tilde{\rho}_N^A(\alpha,v)$ are the nucleon ($N$ = $p$ or $n$)
light-cone momentum and virtuality distributions in nucleus $A$,
$F_2^N(\tilde{x})$ are the free nucleon structure functions,
$\tilde{x}=\frac{Q^2}{2 p\cdot q}$, where and $p$ is the four-momentum of the struck off-shell
nucleon, and the function $f^\text{off}(\tilde{x})$ is the bound nucleon off-shell modification function.
When working in coordinates where $\hat{z}$ is opposite the direction
of $\vec{q}$, $\alpha=A(E+p_z)/m_A$. For brevity we omit the explicit $Q^2$ dependences 
of $F_2^p$, $F_2^n$, and $F_2^A$, but note that they are evaluated at the
same $Q^2$ value.

This convolution combines nuclear wave function input in the form of the nucleon momentum and
virtuality distributions from relatively well-known nuclear structure calculations with partonic input
in the form of nuclear/nucleon structure and off-shell modification functions.   
While the nuclear and free proton structure functions $F_2^A$ and $F_2^p$ are directly measured,
both the neutron-to-proton structure function ratio ${F_2^n} / {F_2^p}$ and bound nucleon off-shell 
modification function $f^\text{off}$ are not. 

As can be seen by Eq.~\ref{Eq:F2_conv}, for a given nucleus $A$,
one can shift strength between ${F_2^n} / {F_2^p}$ and $f^\text{off}$ while still reproducing $F_2^A$ data. Therefore,  ${F_2^n} / {F_2^p}$  and off-shell nucleon modification are correlated when studying a single asymmetric nucleus and/or several symmetric nuclei.

This situation is different for case of $A=3$ mirror nuclei where, 
by exploiting isospin symmetry, i.e. $\tilde{\rho}_p^{^3\text{He}}(\alpha,v) = \tilde{\rho}_n^{^3\text{H}}(\alpha,v)$ 
and $\tilde{\rho}_n^{^3\text{He}}(\alpha,v) = \tilde{\rho}_p^{^3\text{H}}(\alpha,v)$, and by assuming $f^\text{off}$ is the same for both nuclei, 
the convolution equations become
\begin{small}
\begin{equation}
\begin{split}
		&F_2^{^3\text{He}}(x_B) = \\
		& \frac{1}{3}\int_{x_B}^3 \frac{d\alpha}{\alpha} \int_{-\infty}^0 dv \: F_2^p\big(\tilde{x}\big) \left[ 2 \tilde{\rho}_p^{^3\text{He}}(\alpha,v) + 
		 \tilde{\rho}_n^{^3\text{He}}(\alpha,v) \frac{F_2^n(\tilde{x})}{F_2^p(\tilde{x})} \right]   \\
		& \;\;\;\;\;\;\;\;\;\;\;\;\;\;\;\;  \times  \Big(1+v\ f^\text{off}(\tilde{x})\Big), \\
		&F_2^{^3\text{H}}(x_B) = \\
		& \frac{1}{3}\int_{x_B}^3 \frac{d\alpha}{\alpha} \int_{-\infty}^0 dv \: F_2^p\big(\tilde{x}\big) \left[ \tilde{\rho}_n^{^3\text{He}}(\alpha,v) + 
		 2 \tilde{\rho}_p^{^3\text{He}}(\alpha,v) \frac{F_2^n(\tilde{x})}{F_2^p(\tilde{x})} \right]   \\
		& \;\;\;\;\;\;\;\;\;\;\;\;\;\;\;\;  \times  \Big(1+v\ f^\text{off}(\tilde{x})\Big), \\
\label{Eq:F2_A3}
\end{split}
\end{equation}
\end{small}
and the degeneracy between ${F_2^n} / {F_2^p}$ and $f^\text{off}$ is broken.
Therefore, data from mirror nuclei should improve constraints on ${F_2^n} / {F_2^p}$, 
although the exact results obtained can still depend on the assumed model $f^\text{off}$.
Indeed, while previous studies show that inclusive DIS data of deuterium and helium-3 can
be explained using a wide range of functional forms of $f^\text{off}$~\cite{Segarra:2020plg}, 
the MARATHON collaboration's extraction of ${F_2^n} / {F_2^p}$ used only one model for $f^\text{off}$ by KP.

\begin{figure*}[t]
\begin{center}
\includegraphics[width=1.8\columnwidth]{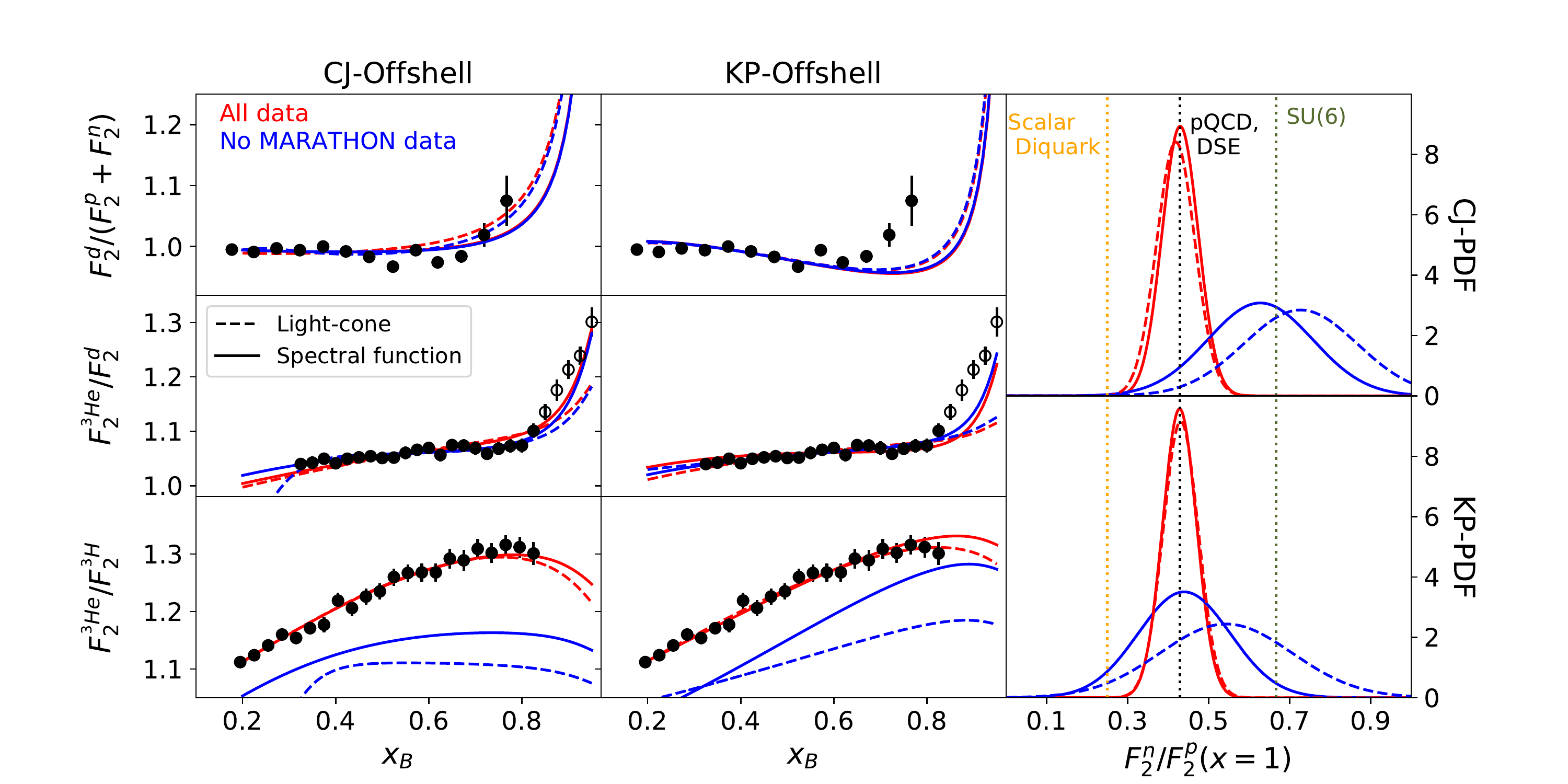}
\caption{ Fit results for the CJ and KP off-shell nucleon modification function. The left and middle columns shows the
          best-fit results compared to the BONuS (top), Seely (middle), and MARATHON (bottom) data for the CJ and KP
          off-shell functions respectively. The right column shows the best-fit result translated into a probability 
          distribution function (PDF) for $F_2^n/F_2^p\big|_{x_B\to1}$, exactly as in Figs. \ref{Fig:res1} and \ref{Fig:res_iso}.
}
\label{Fig:Res_CJ_KP}
\end{center}
\end{figure*}

Here we study the model dependence of the ${F_2^n} / {F_2^p}\big|_{x_B\to1}$ extraction from the MARATHON $^3$He/$^3$H data.
First, we use both light-cone and spectral function approaches for the calculation of $\tilde{\rho}_N^{^3\text{He}}(\alpha,v)$.
Both approaches are detailed in Ref.~\cite{Segarra:2020plg}, but we emphasize that $\tilde{\rho}_N^A(\alpha,v)$ satisfy
baryon and momentum sum rules in the light-cone approach while the momentum sum rule is slightly violated in the spectral function approach. 
Second, we examine the impact of different off-shell modification models.  Note from Eq.~\ref{Eq:F2_A3} that dependence on nucleon 
virtuality is built into the convolution.  
We explore various $\tilde{x}$ dependencies by applying the following set of $f^\text{off}$ functions :
\begin{eqnarray}
	f^\text{off}_\text{const}(\tilde{x}) &=& C, \\
	f^{\text{off},N}_\text{const\;iso}(\tilde{x}) &=& C_N, \\
	f^\text{off}_{\text{lin}\;x}(\tilde{x}) &=&a + b \cdot  \tilde{x}\\
	f^\text{off}_{KP,\;CJ}(\tilde{x}) &=& C(x_0-\tilde{x})(x_1-\tilde{x})(1+x_0-\tilde{x}) 
\end{eqnarray}
The first two models assume modification that is independent of $\tilde{x}$, and is either the same ($f^\text{off}_\text{const}$) 
or different ($f^{\text{off},N}_\text{const\;iso}$) for neutrons and protons.
The third model assumes a linear dependence on $\tilde{x}$. The fourth uses a
more complex function form whose free parameters ($C$, $x_0$, and $x_1$) were
previously determined by the global analyses of Ref.~\cite{KULAGIN2006126} (KP) and~\cite{Accardi:2016qay} (CJ)
and are held fixed in his work.

Using these $f^\text{off}$ parameterizations, we performed a simultaneous fit of Eq.~\ref{Eq:F2_conv} to the ${F_2^d} / ({F_2^p} + {F_2^n})$ 
and ${F_2^{^3\text{He}}} / {F_2^d}$ data by BONuS~\cite{Griffioen:2015hxa} and Seely~\cite{Seely:2009gt} respectively, and examined the
correlation between the off-shell modification parameters and ${F_2^n} / {F_2^p}\big|_{x_B\to1}$.  We then repeated the fits including the
MARATHON ${F_2^{^3\text{He}}} / {F_2^{^3\text{H}}}$ data and examined the impact on the correlations.

In our fits, we used the $\frac{F_2^n(\tilde{x})}{F_2^p(\tilde{x})}$ parametrization of Ref~\cite{Segarra:2020plg}: 
\begin{equation}
\frac{F_2^n(\tilde{x})}{F_2^p(\tilde{x})}\equiv R_{np}(\tilde{x}) = a_{np} \left( 1- \tilde{x}\right)^{b_{np}} + c_{np},
\label{Eq:F2pn}
\end{equation}
which is both functionally flexible and has a single fit parameter $c_{np}$ that describes ${F_2^n} / {F_2^p}\big|_{x_B\to1}$.
This parameterization neglects the weak $Q^2$-dependence of $F_2^n/F_2^p$. We do include the full $Q^2$-dependencee of $F_2^p$,
for which we use the results of the GD11-P global analyis~\cite{GD11P:2011}.

Figure~\ref{Fig:res1} shows the results for fitting the data using the $f^\text{off}_\text{const}$ and $f^\text{off}_{\text{lin}\;x}$ models. 
In all figures, results obtained using light-cone and spectral function formulations of the nuclear wave function are 
shown in dashed and solid lines, respectively. Fits with and without the MARATHON data are shown in red and blue, respectively.
All confidence region contours are shown at the $90\%$ confidence level.

As can be seen, without the MARATHON data a very broad correlation exists between the extracted ${F_2^n} / {F_2^p}\big|_{x_B\to1}$ and the off-shell modification. This results in a wide distribution for ${F_2^n} / {F_2^p}\big|_{x_B\to1}$ that is centered near the SU(6) symmetry prediction of $2/3$. Adding the MARATHON ${F_2^{^3\text{He}}} / {F_2^{^3\text{H}}}$ data significantly limits this correlation and results in ${F_2^n} / {F_2^p}\big|_{x_B\to1}$ = $0.43 \pm 0.04(1\sigma)$ for the constant and linear modification models. These values are consistent with the pQCD and DSE predictions and reject the scalar diquark prediction by $4.5$ standard deviations. In both cases the light-cone and spectral function nuclear wave function formulations have minimal impact on ${F_2^n} / {F_2^p}\big|_{x_B\to1}$.

Next, we examine the case of isospin-dependent modification of protons and neutrons (Fig.~\ref{Fig:res_iso}). In this case, without the MARATHON data ${F_2^n} / {F_2^p}\big|_{x_B\to1}$ is essentially unconstrained. With the MARATHON data, it is it constrained to $0.40 \pm 0.07(1\sigma)$ 
with constant-in-$x$ modification, and to $0.43 \pm 0.10 (1\sigma)$ and $0.49 \pm 0.07 (1\sigma)$ for the SF and LC approaches, respectively,
with linear modification. These constraints 
are in good agreement with the isospin independent results of Fig~\ref{Fig:res1}, but with larger uncertainties that reject the scalar
diquark assumption by only 2--3 standard deviations.

We also show in Fig.~\ref{Fig:Res_CJ_KP} the results of our fit using the CJ and KP $f^\text{off}$ parameterizations. We note again that we did not re-fit their parameters for $f^\text{off}$. While their functional form for $f^\text{off}$ is very different from the simple constant and linear forms discussed above, the resulting constraint on ${F_2^n} / {F_2^p}\big|_{x_B\to1}$ is similar with values of $0.42 \pm 0.04(1\sigma)$ and $0.43 \pm 0.04(1\sigma)$ for CJ and KP respectively.

Looking at the quality of the fits, we note that all models do a very good job describing the 
MARATHON ${F_2^{^3\text{He}}} / {F_2^{^3\text{H}}}$ data. The Seely ${F_2^{^3\text{He}}} / {F_2^d}$ 
data is well reproduced up to $x_B \sim 0.8$, above which the invariant mass, $W$, of the kinematics
drops below $1.4$ GeV.  The situation is similar for the BONuS data, which is well reproduced by all 
calculations except by the KP $f^\text{off}$ parametrization at high-$x_B$, where the kinematics are
also at very low $W$.

Another interesting feature we observe in Figs.~\ref{Fig:res1} and~\ref{Fig:res_iso} is that while the
addition of the MARATHON ${F_2^{^3\text{He}}} / {F_2^{^3\text{H}}}$ data significantly constrains
${F_2^n} / {F_2^p}\big|_{x_B\to1}$, it is less effective in constraining the off-shell modification 
parameters, whose likelihood distributions still span a wide range and have non-negligible sensitivity
to the nuclear wave function formulation (i.e. spectral function vs. light-cone). This reaffirms the
conclusions of Ref.~\cite{Segarra:2020plg} that off-shell modification cannot be well constrained using inclusive DIS data,
even when studying nuclei with different proton-neutron asymmetries.

\begin{figure}[t]
\begin{center}
\includegraphics[width=0.8\columnwidth]{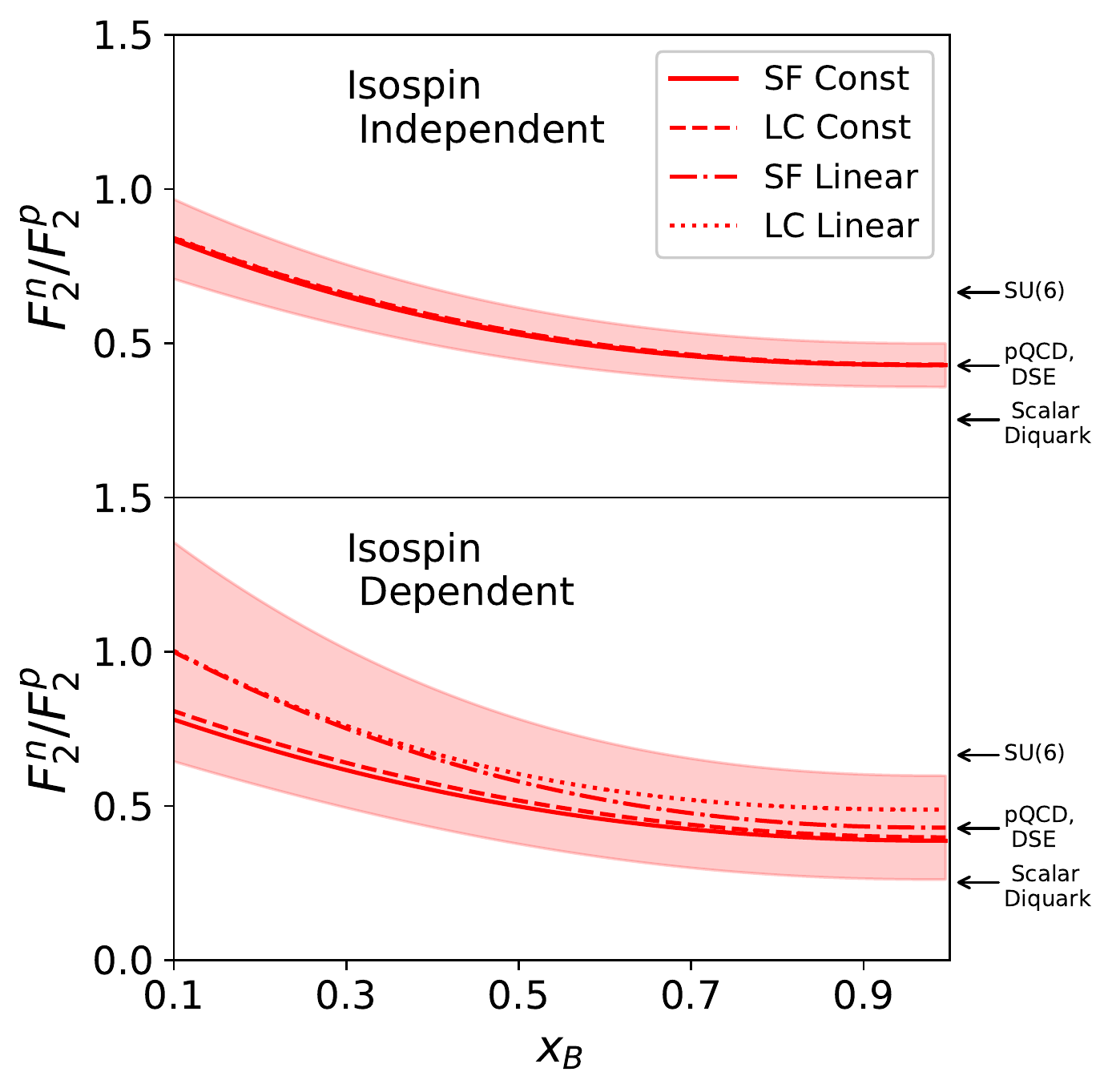}
\caption{ 
	$R_{np}(x_B)$ extracted with MARATHON data from (top) isospin-independent off-shell nucleon modification 
	and (bottom) isospin-dependent off-shell nucleon modification. Constant-in-x and linear-in-x
	off-shell function results are shown with the spectral function and light-cone convolution in different lines.
	The 90\% C.I. band is taken from the trial with the largest uncertainty [(top) linear-in-x, spectral function band 
	is used, (bottom) linear-in-x, light-cone band is used].}
\label{Fig:F2nF2p}
\end{center}
\end{figure}

Last we present the extracted $F_2^n/F_2^p$ ratio as a function of $x_B$ in Fig.~\ref{Fig:F2nF2p} for
both the isospin-independent and isospin-dependent off-shell modifications. As expected, with an isospin-dependent
modification assumption, $F_2^n/F_2^p$ is less constrained, though still consistent with the pQCD and DSE predictions yet
cannot strongly reject SU(6) or scalar diquark predictions. Since we do not currently consider available data constraining $F_2^d$~\cite{ARNEODO1995107} 
at very low-$x_B$, our model uncertainty is large in the low-$x_B$ region ($\sim0.1-0.3$), especially for the isospin-dependent assumption. 
We also note that the $x_B$ value we are sensitive to in $F_2^n/F_2^p$ is smeared out due to the convolution process 
of Eq.~\ref{Eq:F2_conv} (i.e. $\tilde{x} \leq x_B$). However, as seen in Fig.~\ref{Fig:F2nF2p}, the saturation seen around
 $x_B\sim0.7$ gives confidence in our extrapolation to $x_B\rightarrow 1$.

To summarize, we have studied the impact of the MARATHON results on constraints on both the free neutron-to-proton
structure function ratio, $F_2^n/F_2^p$, as $x_B \rightarrow 1$, and the virtuality dependence of the nucleon
structure modification. Whereas MARATHON chose to use the single KP model for the virtuality dependence to extract $F_2^n/F_2^p\big|_{x_B\to1}$,
we have examined the impact of a wide range of possible forms for this unknown dependence. We find that 
for an isospin-independent modification the inclusion of MARATHON data results in a ratio $F_2^n/F_2^p\big|_{x_B\to1} = 0.43 \pm 0.04$, 
regardless of the specific virtuality-dependence. However, for an isospin-dependent modification, the
constraints are less stringent ($F_2^n/F_2^p\big|_{x_B\to1} \approx 0.4 \pm 0.1$) and show variation based
on the specific model used. Furthermore, the inclusion of the MARATHON data does little to constrain
the virtuality-dependence of nucleon modification, and therefore sheds little light on the nature of
the EMC Effect. A wide range of forms and parameters for the nucleon modification function can 
easily accommodate the MARATHON, BONuS, and Seely data. We see this as strong motivation to look
to new observables beyond inclusive DIS to help constrain models of the EMC Effect, such as 
spectator-tagged DIS, on which the results of the BONuS12~\cite{bonus12} and BAND and LAD~\cite{BAND,Segarra:2020txy} experiments are anticipated.

\begin{acknowledgments}
We thank a bunch of people for insightful discussions.
This work was supported by the U.S. Department of Energy, Office of Science, Office of Nuclear Physics under Award Numbers DE-FG02-94ER40818, DE-SC0020240, DE-FG02-96ER-40960, DE-FG02-93ER40771.
\end{acknowledgments}

\bibliography{../../../references.bib}

\end{document}